\documentclass[aps,pra,twocolumn,superscriptaddress]{revtex4-1}

\usepackage{graphicx}
\usepackage{hyperref}
\usepackage{amsmath}
\usepackage{amssymb}
\usepackage{epstopdf}
\usepackage{multirow}
\usepackage{threeparttable}
\usepackage[usenames]{color}

\begin{document}

\title{High-resolution mid-infrared single-photon upconversion ranging}

\author{Shuhong Jiang}
\affiliation{State Key Laboratory of Precision Spectroscopy, East China Normal University, Shanghai 200062, China}

\author{Kun Huang}
\email{khuang@lps.ecnu.edu.cn}
\affiliation{State Key Laboratory of Precision Spectroscopy, East China Normal University, Shanghai 200062, China}
\affiliation{Chongqing Key Laboratory of Precision Optics, Chongqing Institute of East China Normal University, Chongqing 401121, China}
\affiliation{Collaborative Innovation Center of Extreme Optics, Shanxi University, Taiyuan, Shanxi 030006, China}

\author{Tingting Yu}
\affiliation{State Key Laboratory of Precision Spectroscopy, East China Normal University, Shanghai 200062, China}

\author{Jianan Fang}
\affiliation{State Key Laboratory of Precision Spectroscopy, East China Normal University, Shanghai 200062, China}

\author{Ben Sun}
\affiliation{State Key Laboratory of Precision Spectroscopy, East China Normal University, Shanghai 200062, China}

\author{Yan Liang}
\affiliation{School of Optical Electrical and Computer Engineering, University of Shanghai for Science and Technology, Shanghai 200093, China}

\author{Qiang Hao}
\affiliation{School of Optical Electrical and Computer Engineering, University of Shanghai for Science and Technology, Shanghai 200093, China}

\author{E Wu}
\affiliation{State Key Laboratory of Precision Spectroscopy, East China Normal University, Shanghai 200062, China}
\affiliation{Chongqing Key Laboratory of Precision Optics, Chongqing Institute of East China Normal University, Chongqing 401121, China}

\author{Ming Yan}
\affiliation{State Key Laboratory of Precision Spectroscopy, East China Normal University, Shanghai 200062, China}
\affiliation{Chongqing Key Laboratory of Precision Optics, Chongqing Institute of East China Normal University, Chongqing 401121, China}

\author{Heping Zeng}
\email{hpzeng@phy.ecnu.edu.cn}
\affiliation{State Key Laboratory of Precision Spectroscopy, East China Normal University, Shanghai 200062, China}
\affiliation{Chongqing Key Laboratory of Precision Optics, Chongqing Institute of East China Normal University, Chongqing 401121, China}
\affiliation{Shanghai Research Center for Quantum Sciences, Shanghai 201315, China}
\affiliation{Chongqing Institute for Brain and Intelligence, Guangyang Bay Laboratory, Chongqing, 400064, China}

\begin{abstract}
Single-photon laser ranging has widespread applications in remote sensing and target recognition. However, highly-sensitive light detection and ranging (LiDAR) has long been restricted in visible or near-infrared bands. An appealing quest is to extend the operation wavelength into the mid-infrared (MIR) region, which calls for an infrared photon counting system at high detection sensitivity and precise temporal resolution. Here, we devise and demonstrate a MIR upconversion LiDAR based on nonlinear asynchronous optical sampling. Specifically, the infrared probe is interrogated in a nonlinear crystal by a train of pump pulses at a slightly different repetition rate, which favors for a temporal optical scanning at a picosecond timing resolution and a kilohertz refreshing rate over $\sim$50 ns. Moreover, the cross-correlation upconversion trace is temporally stretched by a factor of 2$\times$10$^4$, which can thus be recorded by a low-bandwidth silicon detector. In combination with time-correlated photon-counting technique, the achieved effective resolution is about two orders of magnitude better than the timing jitter of the detector itself, which facilitates a ranging precision of 4 $\mu$m under a low detected flux of 8$\times$10$^{-5}$ photons per pulse. The presented MIR time-of-flight range finder is featured with single-photon sensitivity and high positioning resolution, which would be particularly useful in infrared sensing and imaging in photon-starved scenarios.
\end{abstract}

\maketitle

\section{Introduction}
Laser-based light detection and ranging (LiDAR) has been widely used in various areas of science and technology, including autonomous driving, facial recognition, industrial process monitoring, and space debris tracking \cite{Kim2021NN, Kim2020JOSAB, Steindorfer2020NC, Riemensberger2020Nature}. Among various approaches, time-of-flight (ToF) ranging sets the benchmark for absolute distance metrology, offering high precision, long-range, and fast acquisition \cite{Ye2004OL, Coddington2009NP, Jiang2020NP, Na2020NP, Lee2020Lee}. Moreover, single-photon detection has recently been incorporated into the ToF-based LiDAR system based on the time-correlated single-photon counting (TCSPC) technique \cite{Pawlikowska2017OE, Kong2020OL, Liang2021IEEE}. Such photon-counting laser ranging is particularly pertinent to the low-light-level ranging and imaging context \cite{Sun2016NC, Altmann2018Science, Hadfield2023Optica}, for instance, satellite survey, altimetric measurement, and topographic mapping. Notably, further combination with advanced computational algorithms allows for demonstrating long-distance observation over the terrestrial atmosphere \cite{Li2021Optica} or remote object identification beyond the line of sight \cite{Faccio2020NRP}. However, these state-of-the-art single-photon ranging performances have so far been limited in the visible or near-infrared regions. 

Nowadays, there is a significant impulse to extend the operation wavelength of single-photon LiDAR into the mid-infrared (MIR) region, pertaining to unique features of the reduced photon scattering due to relatively long wavelengths \cite{Flannigan2022JO} and the inherent chemical selectivity based on molecular rovibrational transitions \cite{Schliesser2012NP}. Specifically, MIR LiDAR within transparency windows of the Earth's atmosphere favors remote sensing especially at adverse weather conditions such as mist, fog, and haze \cite{Walsh2016JL}. In virtue of free-space spectroscopy, MIR differential absorption LiDAR (DIAL) systems could facilitate range-resolved investigation of atmospheric gas concentrations \cite{Staehr1985AO}, which holds great potential for pollution surveillance, leakage warning, and meteorological researches \cite{Alvarez2012PNAS, Girard2015AO, Yu2021LSA}. In these envisioned scenarios, sensitive MIR detection is highly demanded to dramatically improve the working distance and/or reduce the transmitting laser power, which however challenges existing infrared detectors based on narrow-bandgap semiconductors \cite{Rogalski2016RPP},  bolometer sensors \cite{Blaikie2019NC}, or emerging low-dimensional materials \cite{Wang2019Small}. Indeed, the attainable detection sensitivity is typically limited to about pW/Hz$^{1/2}$, which is far away from the single-photon level. Recently, superconducting nanowire detectors have been engineered to access MIR photon-counting capability \cite{Marsili2012NL, Chen2021SB}, albeit with consequent trade-off for detection performances and additional system complexity of cryogenic operation. To date, it remains elusive to realize MIR single-photon ranging, which urgently calls for infrared photon-counting systems at high detection sensitivity and high time resolution.

In this context, frequency upconversion detection has been proposed to leverage high-performance silicon detectors by spectrally converting the infrared photons into the visible band, which offers desirable features of high detection efficiency, low dark noise, fast time response, and room-temperature operation \cite{Barh2019AOP, Huang2022NC, Mrejen2020LPR}. Consequently, the single-photon upconversion detector has served as an effective tool in a wide range of infrared applications, such as long-distance quantum communication \cite{Liao2017NP}, quantum-limited optical state characterization \cite{Mancinelli2017NC}, high-sensitivity spectroscopic analysis \cite{Rodrigo2021LPR, Zheng2023LPR, Zhao2023NC}, high-speed optical coherent tomography \cite{Israelsen2019LSA}, and phototoxicity-free biological examination \cite{Junaid2019Optica}. Furthermore, mode selective upconversion detection has recently been configured to facilitate noise-tolerant LiDAR \cite{Shahverdi2018OE, Rehain2020NC}, where only the backscattered signal photons in specific time-frequency modes are efficiently converted due to the intrinsic phase-matching requirement \cite{Ansari2018Optica}. 

Generally, there are two main categories for the upconversion single-photon ranging. One relies on a continuous-wave pumping upconverter, which is suitable for collecting backscattered photons from non-cooperative targets at an unknown distance. The recorded time correlation between the receiving signal and the trigger enables one to identify the absolute distance information \cite{Xia2015OL, Yue2022RS}. But the ranging resolution in the TCSPC configuration is typically limited in the sub-centimeter scale, which is mainly determined by the timing jitter of the detector about tens of picoseconds \cite{Widarsson2020AO, Widarsson2022AO}. An alternative scheme resorts to the pulsed pumping upconverter, where the detected photons are precisely time-stamped by the ultrashort optical pulse, thus significantly improving the axial resolution \cite{Wang2021PRL, Potma2021Optica, Zhang2022PR, Fang2023LSA}. However, the available capture depth is usually confined within tens of centimeters due to a limited travel range of the involved mechanical scanner as the optical delay line \cite{Fang2023LSA, ChenIEEE2023}. Therefore, it necessities the development of novel techniques to address the long-sought-after goal for realizing a high-resolution MIR single-photon LiDAR over a wide operation range.

Here we devise and implement a novel MIR upconversion ranging system based on asynchronous optical sampling, which is featured with single-photon sensitivity, high timing resolution, and wide scanning range. Our realization relies on a dual-comb configuration, where the probe and pump lasers are stabilized at slightly different repetition rates. The reflected MIR probe at 3.1 $\mu$m is asynchronously gated in a nonlinear crystal by the pair-wise ultrashort pump pulse at 1.03 $\mu$m through the sum frequency generation (SFG) process. Consequently, an all-optical temporal scanning is facilitated at a picosecond timing resolution and a kilohertz refreshing rate over a long capture range of 46.45 ns, which results in a time-stretch cross-correlation trace  at the SFG wavelength of 0.77 $\mu$m. The distance information can be real-time monitored from the waveform detected by a low-bandwidth silicon photodiode. Furthermore, MIR single-photon ranging is demonstrated by combing the TCSPC technique and dual-comb metrology, which permits an extremely low detected flux of 8$\times$10$^{-5}$ photons/pulse. Meanwhile, the achieved picosecond temporal resolution is about hundredfold improved over the timing jitter of the used single-photon detector itself, which leads to a high positioning precision of 4 $\mu$m. As a proof-of-principle demonstration, the MIR photon-counting LiDAR is used to retrieve rich information in thickness, reflectivity, and refractive index for a multi-layered silicon wafer under a low-photon-flux illumination. Notably, our approach alleviates the limitation from the intrinsic timing jitter of the photon-counting system, which opens the path toward a sub-$\mu$m ranging precision at the single-photon level. Hence, the presented paradigm for the MIR laser range finder is promising to realize sensitive and precise infrared sensing at a long standoff distance.

We would like to note that the presented approach is inspired from the dual-comb architecture, where the asynchronous pulse trains are used to facilitate the fast and precise optical scanning in the time-of-flight measurement. In the conventional dual-comb scheme, the signal source and local oscillator are phase-locked to generate heterodyne cross-correlation through coherent mode-beating operation between two optical fields \cite{Coddington2009NP}. In contrast, our scheme requires no phase locking for the dual-color sources at disparate wavelengths. The cross-correlation is generated through the nonlinear parametric process, and depends only on the optical field intensity. Moreover, the involved nonlinear optical gating also facilitates sensitive upconversion detection of MIR light. The precise gating of the temporal information and the ability to count infrared photons pave the way to realizing the high-resolution MIR photon-counting LiDAR.
 
\section{Basic principle}
\begin{figure*}[t!]
\includegraphics[width=0.85\textwidth]{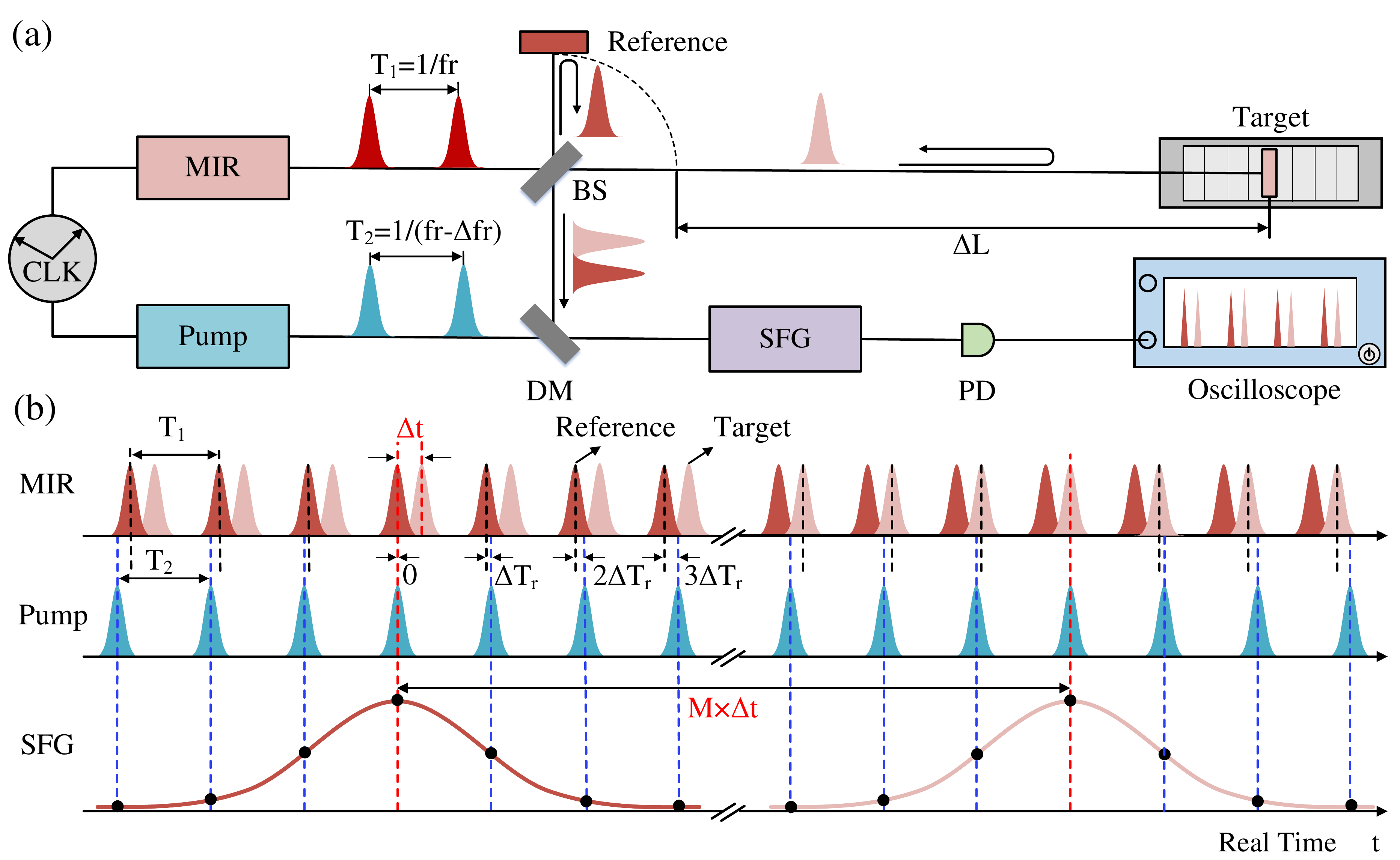}
\caption{Basic concept of MIR distance measurement based on asynchronous upconversion sampling. (a) The MIR probe and the pump beam are from two pulsed lasers, which are stabilized around a repetition rate $f_r$ with a slight frequency difference $\Delta f_r$. The infrared probe is split by a beam splitter (BS) before being steered into reference and target surfaces. The reflections are then spatially mixed with the pump beam via a dichroic mirror (DM). The combined dual-color beams are sent into a nonlinear crystal to perform sum-frequency generation (SFG) in an asynchronous-pumping configuration. The resultant cross-correlation trace is measured by a silicon photodiode (PD), which allows for inferring the distance information through the time of flight. (b) The temporal separation between the signal and reference pulses is measured by using the nonlinear asynchronous optical sampling technique, where the MIR profile is precisely sampled by a pair-wise pump pulse at each period. Consequently, the cross-correlation trace is time stretched with a factor of $M = f_r/\Delta f_r$, which substantially alleviates the requirement for large detection bandwidth or low timing jitter in high-precision measurements.}
\label{fig1}
\end{figure*}

The core process of the proposed MIR laser ranging lies in a SFG-based nonlinear upconversion under asynchronous pulse pumping as illustrated in Fig. \ref{fig1}(a). The involved probe and pump lasers are stabilized at a repetition rate around $f_r$ with a slight difference of $\Delta f_r$. The MIR probe is split and transmitted to a local reference and a distant target. The reflected light is then spatially mixed with the pump within a second-order nonlinear crystal to perform the so-called asynchronous optical sampling \cite{Taschler2023LPR, Dong2018OL, Zhang2014OE}. The resultant cross-correlation trace at the SFG wavelength is finally recorded by a silicon photodiode, which allows for inferring the distance information through the time of flight.

Specifically, the underlying mechanism for the asynchronous upconversion sampling is further detailed in Fig. \ref{fig1}(b). Without loss of generality, the pulse interval for the MIR probe is here assumed to be shorter than that for the pump. As a result, there is a temporal shift for each pair of pulses, given by
\begin{equation}
\Delta T_r = \frac{1}{f_r} - \frac{1}{f_r - \Delta f_r} \approx  \frac{\Delta f_r}{f_r^2}\ .
\label{eq1}
\end{equation}
In the static reference frame of the probe, the pump pulse sweeping at a step of $\Delta T_r$ facilitates an optical temporal scanning, which is an essential feature for dual-comb metrology \cite{Dong2018OL, Zhang2014OE}. The real time for an effective step of $\Delta T_r$ is given by the pulse interval $T = 1/f_r$. Hence, the refreshing period for completing a full scan is derived as
\begin{equation}
T_\text{update} = \frac{T}{\Delta T_r} \times T \approx \frac{1}{\Delta f_r}\ .
\label{eq2}
\end{equation}
It can be seen that $\Delta f_r$ corresponds to the updating rate. Similarly, the scaling factor between the laboratory time and effective time can be determined as
\begin{equation}
M = \frac{T}{\Delta T_r} \approx \frac{f_r}{\Delta f_r} \ .
\label{eq3}
\end{equation}
Through the nonlinear SFG interaction, a cross-correlation intensity profile is generated with a time stretching factor of $M$. Consequently, the time separation $\Delta t$ between the target and reference peaks will be magnified to be $M \times \Delta t$. It is the substantial time magnification that allows us to use a low-bandwidth detector to record the fast time response, which particularly benefits the high-precision MIR measurement where fast and sensitive detectors are currently challenging in this spectral region. Moreover, the stretched waveform also favors to mitigate the limitation of the timing jitter for the single-photon detector in the photon-counting laser ranging scenario \cite{Ren2021IEEE}.

\begin{figure*}[t!]
\includegraphics[width=0.85\textwidth]{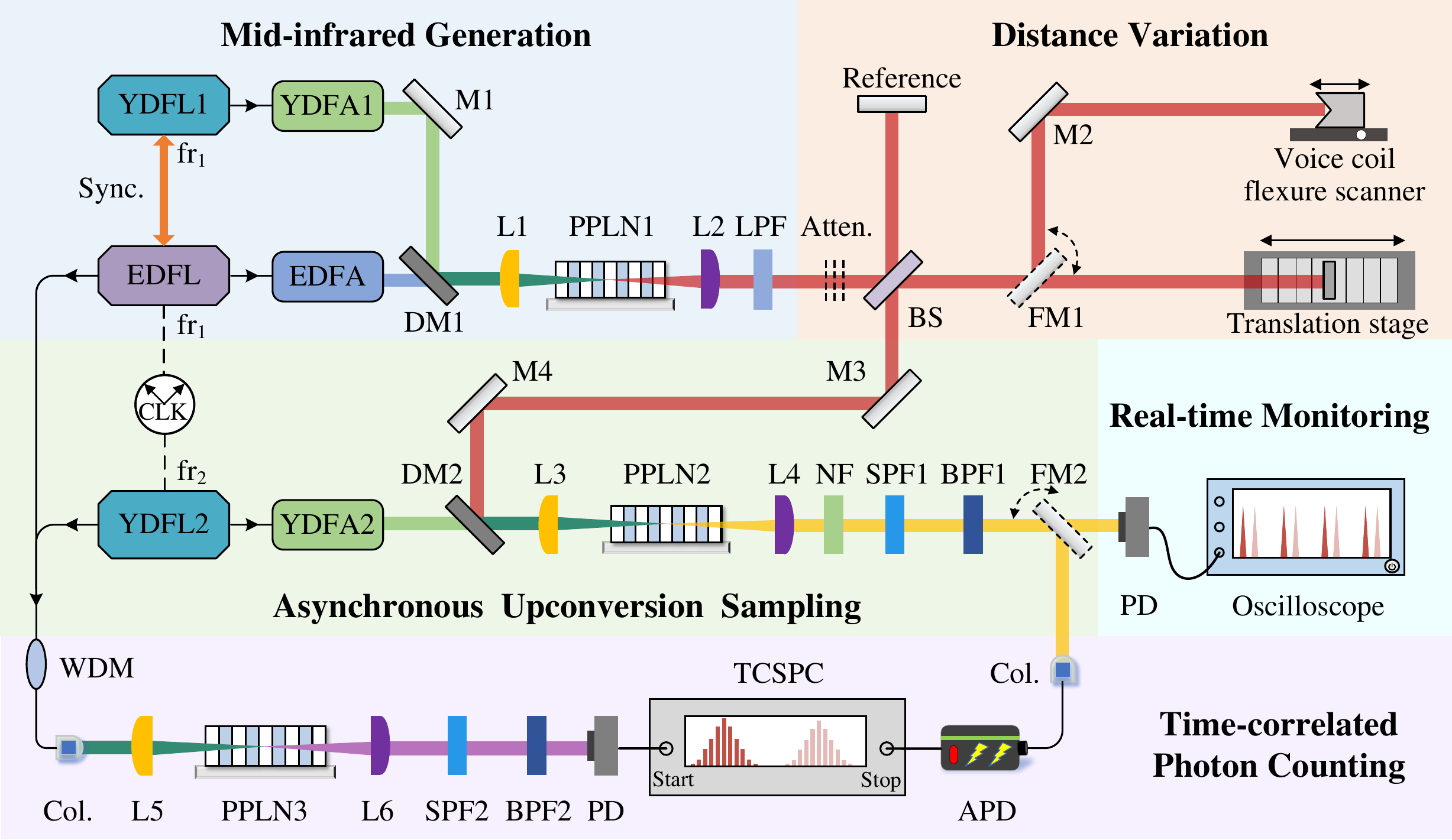}
\caption{Experimental setup for the MIR single-photon upconversion ranging. The involved light sources are prepared from an Er-doped fiber laser (EDFL) and an Yb-doped fiber laser (YDFL2), which are stabilized at slightly different repetition rates $fr_1$ and $fr_2$ by referencing to a rubidium atomic clock (CLK). The output of EDFL at 1.56 $\mu$m is injected into a slave laser (YDFL1) at 1.03 $\mu$m to realize the passive optical synchronization. The synchronized dual-color pulses are injected into a periodical poling lithium niobate (PPLN) crystal to perform the difference frequency generation, which allows for the generation of MIR pulses at 3.1 $\mu$m. The MIR probe is split by a beam splitter (BS), and sent to a local reference and a distant target. The reflected light is then spatially combined with the amplified pump from YDFL2 via a dichroic mirror (DM2). The combined beam is steered into another PPLN crystal to perform an asynchronous upconversion sampling through the sum frequency generation. The asynchronous pulse pumping facilitates a fast and precise optical sampling on the MIR temporal profile, which results in a time-stretch cross-correlation trace at a wavelength of 0.77 $\mu$m. The distance information can be monitored in real time from the measured waveform detected by a low-bandwidth silicon photodiode (PD). Furthermore, MIR ranging at the single-photon level is implemented by using the time-correlated single-photon counting (TCSPC) technique. The start channel is connected to the low-pass-filtered electric pulse from the SFG signal between the EDFL and YDFL2, while the stop channel is connected to the output of a single-photon counting module based on an avalanche photodiode (APD). Finally, the target distance can be measured from the accumulated photon histogram. Yb-doped fiber amplifier (YDFA); Er-doped fiber amplifier (EDFA); L: lens; M: mirror; FM: flip mirror; SPF, LPF, and BPF: short-pass, long-pass, and band-pass filters; NF: notch filter; Atten: optical attenuator; Col: fiber collimator; WDM: wavelength division multiplexer.}
\label{fig2}
\end{figure*}

\section{Experimental Setup}
Figure \ref{fig2} presents the experimental setup for the MIR single-photon LiDAR, which consists of laser source preparation, asynchronous upconversion sampling, and time-correlated counting. The involved light sources originate from an Er-doped fiber laser (EDFL, LangyanTech, ErPico RLocking) at 1.56 $\mu$m and an Yb-doped fiber laser (YDFL2, LangyanTech, YbPico RLocking) at 1.03 $\mu$m. The repetition rate of two mode-locked fiber lasers can be tuned within the range of 21.53$\pm$0.01 MHz based on an intra-cavity piezo-actuator mounted on motorized stage. With the help of a servo system, the repetition rates are stabilized to a reference about $f_r$=21.53 MHz with a difference of $\Delta f_r$=1 kHz. The output of EDFL is injected into a slave laser (YDFL1, LangyanTech, YbPico Elite) to realize an optical synchronization based on the cross-phase modulation effect, which offers a passive fashion to stabilize the relative repetition rate for the dual-color lasers \cite{Zheng2023LPR}. The synchronized pulses are then focused into a periodical poling lithium niobate crystal (PPLN1, with a length of 40 mm and a poling period of 30.3 $\mu$m) to generate the MIR probe at 3.1 $\mu$m through difference frequency generation (DFG). The MIR probe is steered into a reference and a target, and the reflection is spatially mixed with the amplified pump. The combined beams focus into another PPLN crystal (PPLN2, with a length of 10 mm and a poling period of 20.9 $\mu$m) to perform the SFG-based upconversion, where the ultrashort pump serves as a rapid optical gate in the temporal domain. The resulting cross-correlation trace is detected by a silicon photodiode, which facilitates a high-speed ranging measurement. In the low-light situation, the upconversion light is recorded by a single-photon counting module (SPCM, Excelitas, SPCM-AQRH-54-FC), which allows for reconstructing the photon-counting waveform based on a high-precision TCSPC device (Qutools, quTAG). The required timing trigger for the correlation measurement is provided by the SFG pulse that is generated by the two asynchronous lasers of EDFL and YDFL2. In our experiment, the peak conversion efficiency is estimated to be about 1.5$\times 10^{-3}$ at a pump power of 130 mW. The conversion efficiency can be further improved by spectro-temporal engineering of the involved pulses and increasing the pump power via large-mode-area fiber amplifiers \cite{Huang2021PR}. The background noise is measured to be about 5 kHz in the case of blocking the incident infrared signal at the entrance of the upconversion stage, which corresponds to a noise probability about 2.4$\times 10^{-4}$ per pump pulse. The low-noise conversion process is essential to facilitate subsequent ultra-sensitive MIR ranging.

\begin{figure*}[t!]
\includegraphics[width=0.70\textwidth]{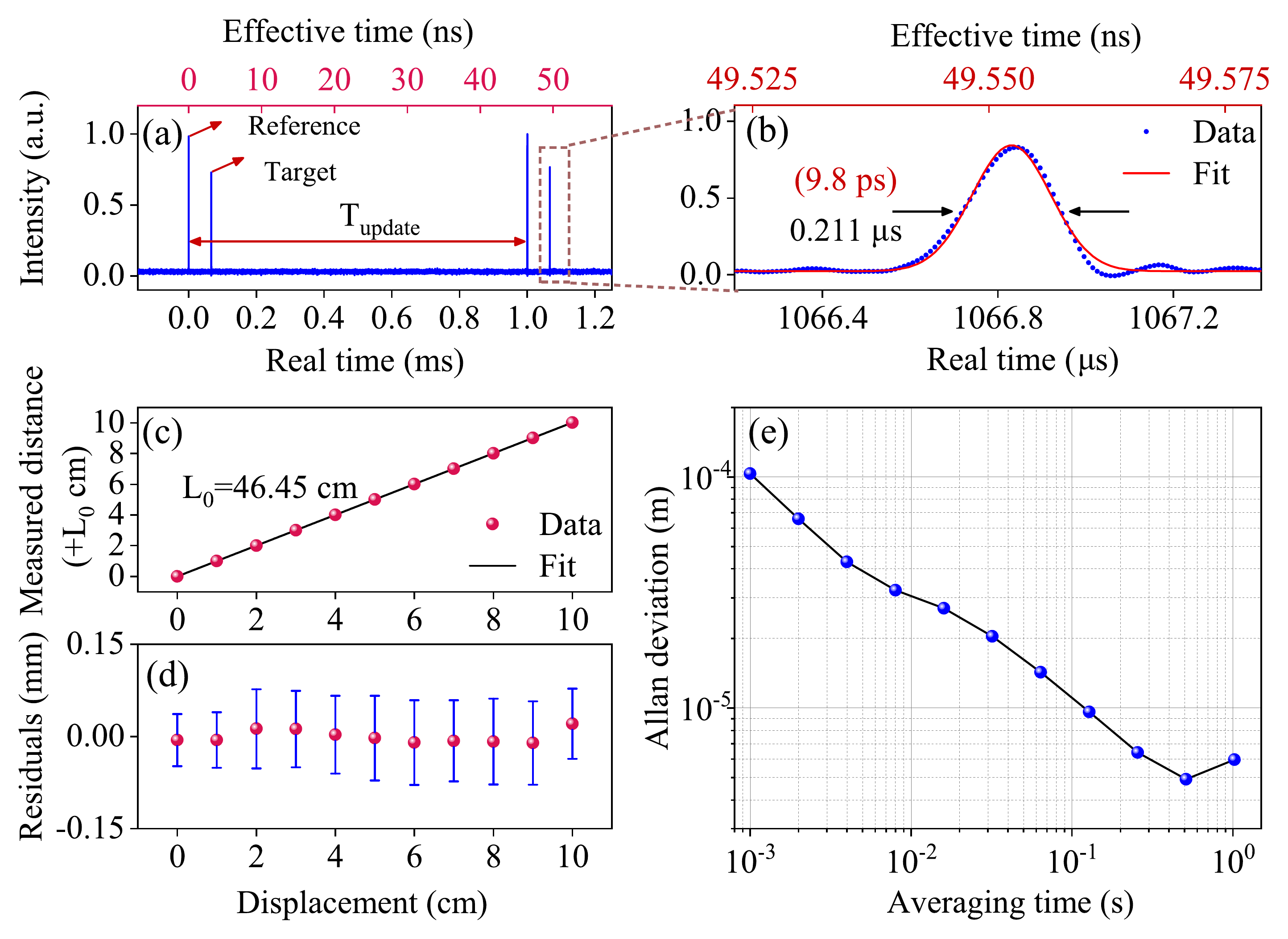}
\caption{High-resolution MIR ranging performance. (a) Recorded time-stretch waveform for the returned infrared probe from the reference and target. The update period $T_\text{update}$ for each measurement is 1 ms, which corresponds to a ranging window about 46.45 ns. (b) Zoom-in of the measured peak for the target. The effective pulse duration of 9.8 ps is dictated by the cross-correlation width between the probe and pump pulses. (c) Measured distance as a function of the displacement for a target mounted on a translational stage. The initial position $L_0$ is 46.45 cm and the travel range is 10 cm. (d) Residual and standard deviation at each measured position. (e) Allan deviation varies with different averaging time at a fixed target distance of $L_0$. The minimum deviation of 5 $\mu$m is achieved for an averaging time about 0.5 s.}
\label{fig3}
\end{figure*}

\section{Results and discussion}
First, the MIR ranging system is characterized to measure the relative displacement of the target. The target is mounted on a linear translation stage (Thorlabs, NRT150/M) with an on-axis accuracy of 2 $\mu$m. The cross-correlation trace is measured by a silicon avalanche detector (Thorlabs, APD410A/M) with a bandwidth of 5 MHz. Figure \ref{fig3}(a) gives the recorded trace at a fixed position, where two adjacent peaks are clearly identified for the reference and target. The scanning step $\Delta T_r$ is calculated to be 2.16 ps, which suffices to sample the cross-correlation trace. The refreshing time $T_\text{update}$ is measured to be 1 ms, which is dictated by the relative repetition rate. The optical sweeping technique allows for a large effective capture range of 46.45 ns, which is inaccessible for a mechanical scanner \cite{ChenIEEE2023}. The full cover of the pulse period of the probe is essential to collect signals from any distance, thus eliminating the dead zone for the ranging. The cross-correlation peak for the target is zoomed in Fig. \ref{fig3}(b), which indicates a pulse duration of 0.211 $\mu$s in laboratory time. The effective duration is inferred to be 9.8 ps by using the conversion factor $M = 2.153 \times 10^4$, which agrees with the width of the convolution between the probe and pump pulses. The time-stretch operation greatly relaxes the bandwidth requirement for fast signal detection and processing.

For distance measurements, the SFG intensity profile for each peak is fitted with a Gaussian function, which allows for precise identification of the peak positions. The measured time separation $\Delta T$ between the target and reference peaks can be converted to the distance according to $\Delta L = c \Delta T / (2 M n_g)$, where $c$ is the speed of light in vacuum, $n_g$ is the group refractive index of air. Here, the distance calculation is relatively simpler than that in conventional dual-comb metrology. Indeed, the distance measurement based on the temporal interferograms typically requires either Fourier transform for phase demodulation or Hilbert transform for carrier frequency elimination \cite{Kim2020JOSAB, Ren2021IEEE}. In contrast, the presented nonlinear optical sampling only depends on the optical field intensity. Such an incoherent operation not only enhances the measurement robustness with a resilience to phase variations, but also simplifies the laser preparation without the need to stabilize the carrier-envelop phase \cite{Dong2018OL, Zhang2014OE}.

\begin{figure*}[t!]
\includegraphics[width=0.75 \textwidth]{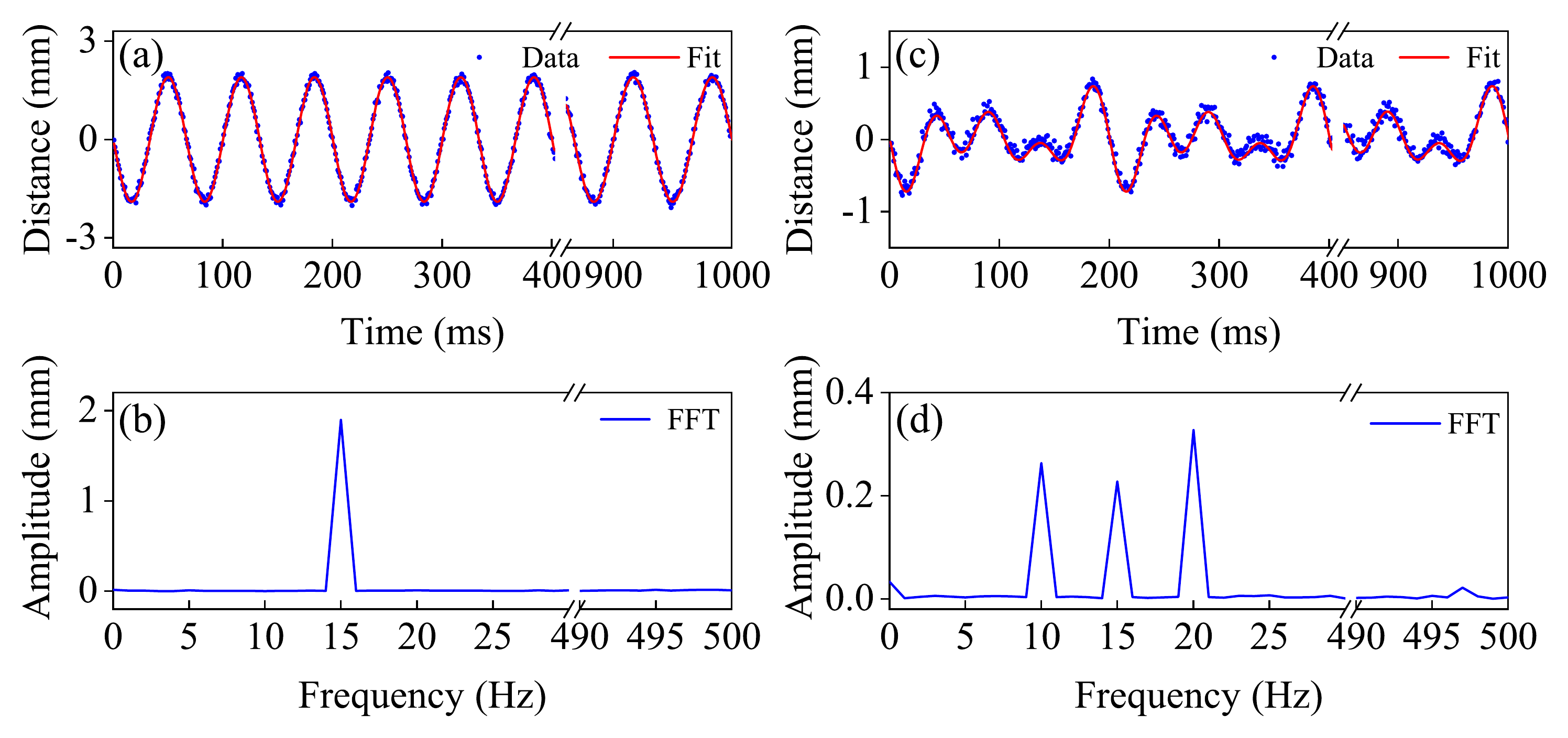}
\caption{High-speed MIR ranging for a rapidly moving voice-coil actuator. (a) Measured oscillating trace at a frequency of 15 Hz. (b) Involved frequency component is revealed by using the fast Fourier transform (FFT). (c) Recorded temporal trace when the actuator is driven by a synthesized voltage. (d) Multiple frequencies are identified via the Fourier analysis.}
\label{fig4}
\end{figure*}

Figure \ref{fig3}(c) presents the measured distance as a function of the target displacement. The linear dependence indicates a high measurement accuracy, which is manifested in the corresponding residuals shown in Fig. \ref{fig3}(d). The ranging uncertainty is about 100 $\mu$m for a single measurement, which can be decreased to 5 $\mu$m for a longer averaging time as seen from the Allan deviation in Fig. \ref{fig3}(e). In our proof-of-principle demonstration, the target is placed at an original distance $L_0$ = 46.45 cm. Actually, the non-ambiguous range for the current experimental setting is about 6.97 m defined by $c/(2 f_r n_g)$. Similar to dual-comb ranging, the operation distance can be significantly extended by tuning the repetition rate for one laser source \cite{Ye2004OL, Zhang2014OE}, or simply switching the repetition rates of the two combs in a separate measurement \cite{Kim2020JOSAB, Coddington2009NP}. In the latter case, the non-ambiguous range can reach to 150 km, which is set by the laboratory dual-comb period of $1/\Delta f_r$=1 ms. The underlying challenge for the long-distance ranging lies in collecting sufficient photons from remote target, which merits the use of sensitive range finders as demonstrated in this work. Furthermore, the combination of advanced computational techniques would enhance the signal-to-noise ratio (SNR) in photon-starved scenarios \cite{Li2021Optica}.

\begin{figure*}[t!]
\includegraphics[width=0.75\textwidth]{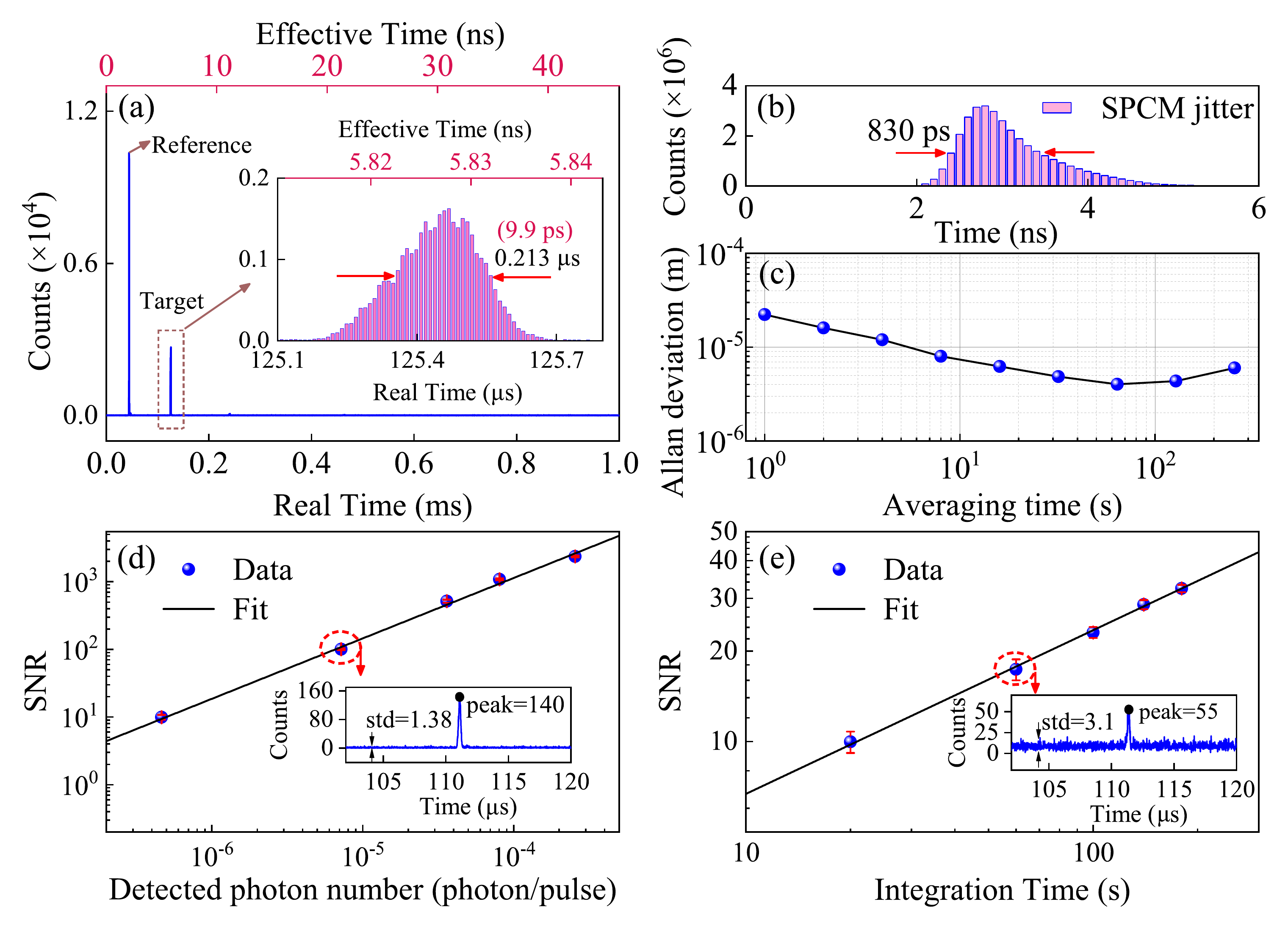}
\caption{High-sensitivity MIR ranging based on the time-correlated photon-counting technique. (a) Measured coincidence histogram for the returned infrared photons. Inset shows the zoom-in for recorded target peak. The peak width is about 0.213 $\mu$s, which corresponds to an effective time resolution of 9.9 ps. (b) Measured timing jitter of the used single-photon detector, which is the main limiting factor for ranging resolution in conventional ToF schemes. (c) Allan deviation measured at a detected optical energy of 8$\times$10$^{-5}$ photons/pulse. The minimum deviation of 4 $\mu$m is achieved for an averaging time about 64 s. (d,e) Signal-to-noise ratio as a function of the detected power (d) and integration time (e). Insets show the measured histograms at representative conditions.}
\label{fig5}
\end{figure*}

Now we turn to demonstrate the high-speed MIR ranging performance. An aluminum retroreflector as the target is mounted at a voice coil flexure scanner (Thorlabs, VCFL35). The scanner is driven by an arbitrary waveform generator (Rigol, DG4162). Figure \ref{fig4}(a) presents the reconstructed trajectory for the rapidly oscillating scanner, which is well fitted by a sinusoidal function. The 1-kHz sampling rate is sufficient to observe the dynamic position change. After a fast Fourier transform (FFT), the peak in the spectral domain clearly indicates the driven frequency of 15 Hz, as shown in Fig. \ref{fig4}(b). Furthermore, a more complicated trace in Fig. \ref{fig4}(c) can be generated by driving the scanner with a synthesized waveform. The corresponding FFT spectrum correctly reveals the frequencies and amplitudes for the three sinusoidal components set in the experiment. In contrast to mechanical scanners, the optical temporal sweeping is inertia-free, which thus favors rapid ToF distance measurements for a moving target. The refreshing rate can further be boosted by increasing the repetition rate difference, while a higher repetition rate itself should be adapted to retain a suitable sampling step \cite{Kim2020JOSAB, Ren2021IEEE}.

\begin{figure}[b!]
\includegraphics[width=0.95\columnwidth]{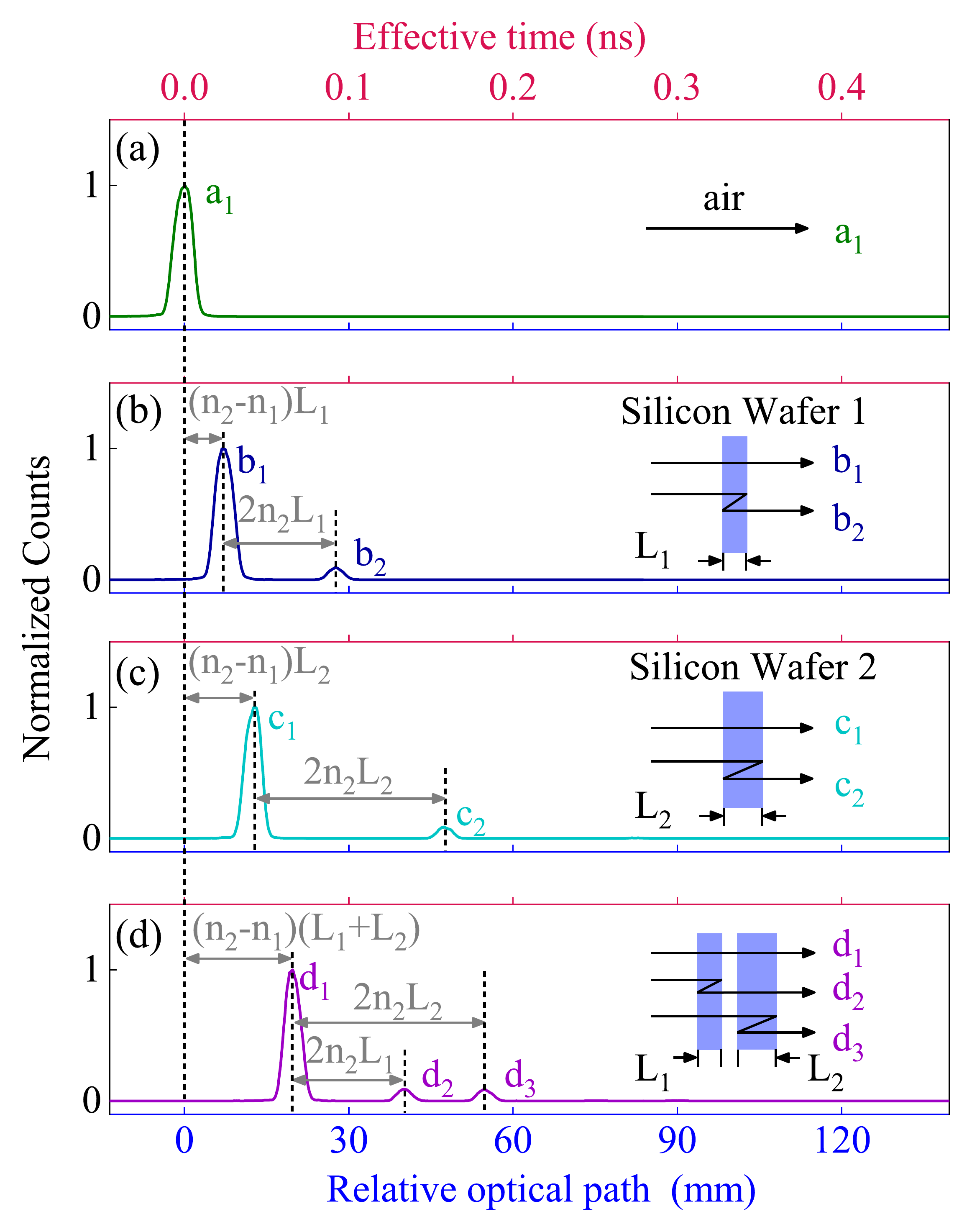}
\caption{Photon-counting MIR depth measurement for multi-interface structures. (a) Measured reference histogram in the absence of samples. (b-d) Measured histograms in the presence of silicon wafers with a thickness of $L_1$ (b), a thickness of $L_2$ (c), and both (d). The involved optical paths are indicated for the observed peaks in each case.}
\label{fig6}
\end{figure}

In the following, we proceed to investigate the high-sensitivity competence of the MIR photon-counting ranging. The implemented parametric upconversion detection leverages silicon-based photon counter with high detection sensitivity and fast time response, which hence provides superior performances beyond that for currently available MIR detectors at room temperature \cite{Hadfield2023Optica}. To emulate the photon-starved scenario, the MIR probe is attenuated by a series of calibrated neutral density filters such that the detected photon number per pulse is much less than one \cite{Kong2020OL, Shahverdi2018OE}. Accordingly, the detector is replaced with a silicon-based SPCM. The detected events are connected to a stop channel of a TCSPC device, while the start channel is triggered by the generated pulses from the SFG between the YDFL2 and EDFL, as depicted in Fig. \ref{fig2}. The recorded time correlation between the receiving signal and the trigger enables one to identify the absolute distance information. The timing jitter in root mean square of the TCSPC is specified to be below 6.4 ps. The bin width and the interrogation window are configured to be 100 ns and 1 ms, respectively. The MIR illumination energy is attenuated to be about 0.8 fJ/pulse. The detected photon number at the SPCM is estimated to be 8$\times$10$^{-5}$ per pulse after taking into account the duty cycle of 2.1$\times$10$^{-4}$ for the asynchronous sampling and the total detection efficiency of 3.2$\times$10$^{-5}$ for the upconversion detector. The total detection efficiency is limited by several factors in the experiment. The main loss is ascribed to the collection efficiency of 10\% for the retro-reflected radiation from the target and the internal conversion efficiency of 0.15\% within the nonlinear crystal. Additionally, the spectral filtering stage has a power transmission of 65\% for the upconverted signal. The SFG signal is coupled into a single-mode fiber with an efficiency of 50\%, before being recorded by a silicon photon counter with a detection efficiency of 65\%. The total detection efficiency could be further improved with a synergic optimization on each stage from light collection, nonlinear conversion, spectral filtering to photon recording \cite{Huang2021PR}. The total acquisition time is set to be about 10 s for collecting sufficient photons to build a high-contrast histogram. The recorded coincidence histogram is presented in Fig. \ref{fig5}(a), which again shows two prominent peaks for the reference and target. A zoom-in illustration of the target peak is given in inset. The histogram width is identical to the one obtained for analogue optical detection. Notably, the effective time resolution of 9.9 ps is achieved, which is much smaller than the 830-ps timing jitter of the SPCM, as measured from the histogram shown in Fig. \ref{fig5}(b).

The ranging precision at the single-photon level is illustrated in the Allan deviation trace shown in Fig. \ref{fig5}(c). At an average time of 64 s, the precision reaches to 4 $\mu$m, which is significantly smaller than previous demonstrations \cite{Widarsson2020AO, Widarsson2022AO}. The presented time-stretch correlated counting technique overcomes the intrinsic timing uncertainty of single-photon detectors \cite{Yang2022SB}, which hence allows for obtaining a precision beyond the reach of conventional ToF single-photon ranging systems \cite{Liang2021IEEE, Li2021Optica}. Furthermore, the SNR for the recorded histogram is investigated as a function of the detected power and the integration time, as shown in Figs. \ref{fig5}(d) and (e), respectively. The SNR is defined as the ratio between the peak amplitude and the standard deviation of the noise floor. The SNR is particularly important for single-photon ranging since the collected signal is extremely limited. A high SNR usually leads to a better precision. As expected, a stronger illumination power or a longer accumulation time enables one to collect more photons, which thus improves the SNR. Notably, tremendous effort has been devoted to developing advanced algorithms for realizing photon-efficient ranging or imaging, with an aim to achieve a high SNR with a small number of photons \cite{Altmann2018Science, Hadfield2023Optica}.

In comparison to visible or near-infrared wavelengths, the MIR light allows for penetrating through semiconductor materials. For instance, the silicon and germanium substrates are opaque for the optical spectra below 1.1 and 2.0 $\mu$m, respectively. As a proof-of-principle demonstration, we finally use the MIR single-photon ranging system to perform a non-invasive investigation on stacked silicon wafers as a multiple-layer object. The back-scattered photons from the interior interfaces provide rich information on the reflectivity and depth of the imbedded surfaces, as well as the refractive index of the medium \cite{Zhang2022PR, Fang2023LSA}. A reference histogram in Fig. \ref{fig6}(a) is measured in the absence of the sample, where the peak position is defined as the origin. Because the refractive index of the silicon $n_2$ is relatively higher than that for the air $n_1$, the histogram peak in Fig. \ref{fig6}(b) is delayed after inserting a silicon wafer with a thickness of 3 mm. Specifically, two prominent peaks are observed, which correspond to the optical-path configurations as presented in inset. From the measured relative distances, the geometric thickness and refractive index are inferred to be 2.94 mm and 3.49, respectively. Moreover, the power reflectivity at each surface is measured to be 0.304, which is consistent to the normal-incidence Fresnel reflection given by $(n_2-n_1)^2/(n_2+n_1)^2$. At the presence of a 5-mm-thickness wafer, the recorded peaks move at a larger delay, as illustrated in Fig. \ref{fig6}(c). As shown in Fig. \ref{fig6}(d), more peaks emerge with both inserted wafers, thus showing a depth-resolvable capability for multi-layer structures. Pertinent to the transparency window for semiconductors and polymer materials \cite{Israelsen2019LSA, Potma2021Optica}, the MIR sensitive and precise depth ranging would open an effective way to perform non-destructive defect inspection or stand-off surface profiling in extreme scenarios for industrial quality control and process monitoring.

\section{Conclusion}
Although single-photon laser ranging is widely demonstrated in a variety of applications, yet the operation wavelength has long been restricted in the visible or near-infrared region due to the availability of sensitive and fast optical detectors \cite{Altmann2018Science, Hadfield2023Optica}. Our work addresses the long-standing quest to extend the operation wavelength of photon-counting ranging into the MIR regime. The presented approach of the asynchronous upconversion sampling technique not only facilitates the optical temporal scanning at high speed and high precision, but also enables the sensitive upconversion detection for the gated infrared signals with a silicon-based visible photodiode operating at room temperature. The precise gating of the temporal information and the ability to count infrared photons constitute the key to the first realization of high-resolution MIR photon-counting LiDAR under low-light-level illumination. 

In comparison to MIR upconversion LiDAR systems based on continuous-wave pumping \cite{Xia2015OL, Yue2022RS, Widarsson2020AO, Widarsson2022AO}, the involved pulsed pumping configuration favors the improvement of nonlinear conversion efficiency due to the high peak intensity and the reduction of background noise within a narrow time window. Meanwhile, the asynchronous pumping operation allows a long sweeping range while maintaining the ultrashort temporal resolution, which overcomes the limited scanning range of the mechanical delay line in the coincidence-pumping scheme \cite{Wang2021PRL, ChenIEEE2023, Huang2021PR}. Another notable feature for the presented approach is the time-stretch cross-correlation trace, which enables us to achieve a photon-counting temporal resolution much smaller than the timing jitter of the used single-photon detectors \cite{Yue2022RS, Widarsson2020AO, Widarsson2022AO}. This unique feature significantly alleviates the bandwidth requirement for fast detection and processing  as typically required in traditional TCSPC architecture. We note that one remaining challenge for the presented photon-counting MIR ranging system is to shorten the acquisition time for observing dynamic scenes in the low-light level. The adopt of computational techniques may facilitate a photon-efficient LiDAR, where the target can be identified from a histogram with much fewer photons \cite{Li2021Optica, Altmann2018Science}.

In conclusion, our work establishes an effective path to achieving high-performance MIR laser LiDAR with single-photon sensitivity, picosecond timing resolution, and large non-ambiguous range. The ranging resolution can be further enhanced to the nanometer level by using shorter optical pulses \cite{Kim2020JOSAB, Coddington2009NP, Lee2020Lee}, without suffering from the timing-jitter restriction of the photon detection and counting devices. Moreover, enhanced performances with a higher refreshing rate and a finer sweeping step can be obtained by properly increasing the repetition rates and their difference for the dual-color lasers \cite{Ren2021IEEE}. Notably, the involved asynchronous laser sources could be substantially simplified by resorting to recent advances of single-cavity dual combs \cite{Yang2023FP}. In addition, it is feasible to extend the presented approach into longer-infrared or terahertz regions \cite{Rodrigo2021LPR}, where high-sensitivity and high-resolution distance measurements are highly demanded. We believe that the implemented MIR single-photon ranging system would promote a variety of low-light-level applications in remote sensing, environmental monitoring, meteorological observation, and defense surveillance.
\newline

\section*{Acknowledgements}
This work was supported by National Key Research and Development Program (2021YFB2801100), National Natural Science Foundation of China (62175064, 62235019, 12022411, 62035005); Shanghai Pilot Program for Basic Research (TQ20220104); Natural Science Foundation of Chongqing (CSTB2023NSCQ-JQX0011, CSTB2022NSCQ-MSX0451, CSTB2022NSCQ-JQX0016); Shanghai Municipal Science and Technology Major Project (2019SHZDZX01); Fundamental Research Funds for the Central Universities.

\section*{Conflict of Interest}
The authors declare no conflict of interests.

\section*{Data Availability Statement}
The data that support the findings of this study are available from the corresponding author upon reasonable request.

\section*{Keywords}
mid-infrared ranging, single-photon ranging, frequency upconversion, nonlinear optical sampling


\begin{thebibliography}{100}

\bibitem{Kim2021NN} I. Kim, R. J. Martins, J. Jang, T. Badloe, S. Khadir, H.-Y. Jung, H. Kim, J. Kim, P. Genevet, and J. Rho,  ``Nanophotonics for light detection and ranging technology," Nat. Nanotechnol. \textbf{16}, 508-524 (2021).

\bibitem{Kim2020JOSAB} W. Kim, J. Jang, S. Han, S. Kim, J. S. Oh, B. S. Kim, Y.-J. Kim, and S.-W. Kim, ``Absolute laser ranging by time-of-flight measurement of ultrashort light pulses [Invited]," J. Opt. Soc. Am. \textbf{37}, B27-B35 (2020).

\bibitem{Steindorfer2020NC} M. A. Steindorfer, G. Kirchner, F. Koidl, P. Wang, B. Jilete, and T. Flohrer, ``Daylight space debris laser ranging," Nat. Commun. \textbf{11}, 3735 (2020). 
  
\bibitem{Riemensberger2020Nature} J. Riemensberger, A. Lukashchuk, M. Karpov, W. Weng, E. Lucas, J. Liu, and T. J. Kippenberg, ``Massively parallel coherent laser ranging using a soliton microcomb," Nature \textbf{581}, 164-170 (2020).

\bibitem{Ye2004OL} J. Ye, Absolute measurement of a long, arbitrary distance to less than an optical fringe," Opt. Lett. \textbf{29}, 1153-1155 (2004).

\bibitem{Coddington2009NP} I. Coddington, W. C. Swann, L. Nenadovic, and N. R. Newbury, ``Rapid and precise absolute distance measurements at long range," Nat.  Photonics \textbf{3}, 351-356 (2009).

\bibitem{Jiang2020NP} Y. Jiang, S. Karpf, and B. Jalali, ``Time-stretch LiDAR as a spectrally scanned time-of-flight ranging camera," Nat. Photonics \textbf{14}, 14-18 (2020).
  
\bibitem{Na2020NP} Y. Na, C.-G. Jeon, C. Ahn, M. Hyun, D. Kwon, J. Shin, and J. Kim, ``Ultrafast, sub-nanometre-precision and multifunctional time-of-flight detection," Nat. Photonics \textbf{14}, 355-360 (2020).
  
\bibitem{Lee2020Lee} J. Lee, Y.-J. Kim, K. Lee, S. Lee, and S.-W. Kim, ``Time-of-flight measurement with femtosecond light pulses," Nat. Photonics \textbf{4}, 716-720 (2010).

\bibitem{Pawlikowska2017OE} A. M. Pawlikowska, A. Halimi, R. A. Lamb, and G. S. Buller, ``Single-photon three-dimensional imaging at up to 10 kilometers range," Opt. Express \textbf{25}, 11919-11931 (2017).

\bibitem{Kong2020OL} L. Kong, Q. Zhao, K. Zheng, H. Lu, S. Chen, X. Tao, H. Wang, H. Hao, C. Wan, X. Tu, L. Zhang, X. Jia, L. Kang, J. Chen, and P. Wu, ``Noise-tolerant single-photon imaging with a superconducting nanowire camera," Opt. Lett. \textbf{45}, 6732-6735 (2020).
  
\bibitem{Liang2021IEEE} Y. Liang, B. Xu, Q. Fei, W. Wu, X. Shan, K. Huang, and H. Zeng, ``Low-Timing-Jitter GHz-Gated InGaAsInP Single-Photon Avalanche Photodiode for LIDAR," IEEE  J. Sel .Top. Quantum Electron. \textbf{28}, 3801807 (2021).

\bibitem{Sun2016NC} M.-J. Sun, M. P. Edgar, G. M. Gibson, B. Sun, N. Radwell, R. Lamb, and M. J. Padgett, ``Single-pixel three-dimensional imaging with time-based depth resolution," Nat. Commun. \textbf{7}, 12010 (2016).

\bibitem{Altmann2018Science} Y. Altmann, S. McLaughlin, M. J. Padgett, V. K. Goyal, A. O. Hero, and D. Faccio, ``Quantum-inspired computational imaging," Science \textbf{361}, eaat2298 (2018).

\bibitem{Hadfield2023Optica} R. H. Hadfield, J. Leach, F. Fleming, D. J. Paul, C. H. Tan, J. S. Ng, R. K. Henderson, and G. S. Buller, ``Single-photon detection for long-range imaging and sensing," Optica \textbf{10}, 1124-1141 (2023).

\bibitem{Li2021Optica} Z.-P. Li, J.-T. Ye, X. Huang, P.-Y. Jiang, Y. Cao, Y. Hong, C. Yu, J. Zhang, Q. Zhang, C.-Z. Peng, F. Xu, and J.-W. Pan, ``Single-photon imaging over 200 km," Optica \textbf{8}, 344-349 (2021).

\bibitem{Faccio2020NRP} D. Faccio, A. Velten, and G. Wetzstein, ``Non-line-of-sight imaging," Nat. Rev. Phys. \textbf{2}, 318-327 (2020).

\bibitem{Flannigan2022JO} L. Flannigan, L. Yoell, and C. Xu, ``Mid-wave and long-wave infrared transmitters and detectors for optical satellite communications-a review," J. Opt. \textbf{24}, 043002 (2022).
  
\bibitem{Schliesser2012NP} A. Schliesser, N. Picqu\'{e}, and T. W. H\"{a}nsch, ``Mid-infrared frequency combs," Nat. Photonics \textbf{6}, 440-449 (2012).

\bibitem{Walsh2016JL} B. M. Walsh, H. R. Lee, and N. P. Barnes, ``Mid infrared lasers for remote sensing applications," J. Lumin. \textbf{169}, 400-405 (2016).

\bibitem{Staehr1985AO}  W. Staehr, W. Lahmann, and C. Weitkamp, ``Range-resolved differential absorption lidar: optimization of range and sensitivity," Appl. Opt. \textbf{24}, 1950-1956 (1985)

\bibitem{Alvarez2012PNAS} R. A. Alvarez, S. W. Pacala, J. J. Winebrake, W. L. Chameides, and S. P. Hamburg, ``Greater focus needed on methane leakage from natural gas infrastructure," Proc. Natl. Acad. Sci. USA \textbf{109}, 6435-6440 (2012).

\bibitem{Girard2015AO} S. Lambert-Girard, M. Allard, M. Pich\'{e}, and F. Babin, ``Differential optical absorption spectroscopy lidar for mid-infrared gaseous measurements," Appl. Opt. \textbf{54}, 1647-1656 (2015).

\bibitem{Yu2021LSA} S. Yu, Z. Zhang, H. Xia, X. Dou, T. Wu, Y. Hu, M. Li, M. Shangguan, T. Wei, L. Zhao, L. Wang, P. Jiang, C. Zhang, L. You, L. Tao, and J. Qiu, ``Photon-counting distributed free-space spectroscopy," Light  Sci. Appl. \textbf{10}, 212 (2021).

\bibitem{Rogalski2016RPP} A. Rogalski, P. Martyniuk, and M. Kopytko,  ``Challenges of small-pixel infrared detectors: a review," Rep. Prog. Phys. \textbf{79}, 046501 (2016).

\bibitem{Blaikie2019NC} A. Blaikie, D. Miller, and  B. J. Alem\'{a}n, ``A fast and sensitive room-temperature graphene nanomechanical bolometer," Nat. Commun. \textbf{10}, 4726 (2019). 

\bibitem{Wang2019Small} P. Wang, H. Xia, Q. Li, F. Wang, L. Zhang, T. Li, P. Martyniuk, A. Rogalski, and W. Hu, ``Sensing Infrared Photons at Room Temperature: From Bulk Materials to Atomic Layers," Small \textbf{15}, 1904396 (2019).

\bibitem{Marsili2012NL} F. Marsili, F. Bellei, F. Najafi, A. E. Dane, E. A. Dauler, R. J. Molnar, and K. K. Berggren, ``Efficient Single Photon Detection from 500 nm to 5 $\mu$m Wavelength," Nano. Letters  \textbf{12}, 4799-4804 (2012).

\bibitem{Chen2021SB} Q. Chen, R. Ge, L. Zhang, F. Li, B. Zhang, F. Jin, H. Han, Y. Dai, G. He, Y. Fei, X. Wang, H. Wang, X. Jia, Q. Zhao, X. Tu, L. Kang, J. Chen, and P. Wu, ``Mid-infrared single photon detector with superconductor Mo$_{80}$Si$_{20}$ nanowire," Sci. Bull. \textbf{66}, 965-968 (2021).
  
\bibitem{Barh2019AOP} A. Barh, P. J. Rodrigo, L. Meng, C. Pedersen, and P. Tidemand-Lichtenberg, ``Parametric upconversion imaging and its applications," Adv. Opt. Photonics \textbf{11}, 952-1019 (2019).

\bibitem{Huang2022NC} K. Huang, J. Fang, M. Yan, E Wu, and H. Zeng, ``Wide-field mid-infrared single-photon upconversion imaging," Nat. Commun. \textbf{13}, 1077 (2022).

\bibitem{Mrejen2020LPR} M. Mrejen, Y. Erlich, A. Levanon, and H. Suchowski, ``Multicolor Time-Resolved Upconversion Imaging by Adiabatic Sum Frequency Conversion," Laser Photonics Rev. \textbf{14}, 2000040 (2020).

\bibitem{Liao2017NP} S.-K. Liao, H.-L. Yong, C. Liu, G.-L. Shentu, D.-D. Li, J. Lin, H. Dai, S.-Q. Zhao, B. Li, J.-Y. Guan, W. Chen, Y.-H. Gong, Y. Li, Z.-H. Lin, G.-S. Pan, J. S. Pelc, M. M. Fejer, W.-Z. Zhang, W.-Y. Liu, J. Yin, J.-G. Ren, X.-B. Wang, Q. Zhang, C.-Z. Peng, and J.-W. Pan, ``Long-distance free-space quantum key distribution in daylight towards inter-satellite communication," Nat.  Photonics \textbf{11}, 509-513 (2017).

\bibitem{Mancinelli2017NC} M. Mancinelli, A. Trenti, S. Piccione, G. Fontana, J. S. Dam, P. Tidemand-Lichtenberg, C. Pedersen, and L. Pavesi, ``Mid-infrared coincidence measurements on twin photons at room temperature," Nat.  Commun. \textbf{8}, 15184 (2017).

\bibitem{Rodrigo2021LPR} P. J. Rodrigo, L. H$\o$gstedt, S.  M.  M.  Friis, L. R. Lindvold, P. Tidemand-Lichtenberg, and C. Pedersen, ``Room-Temperature, High-SNR Upconversion Spectrometer in the 6-12 $\mu$m Region," Laser Photonics Rev. \textbf{15}, 2000443 (2021).
  
\bibitem{Zheng2023LPR} T. Zheng, K. Huang, B. Sun, J. Fang, Y. Chu, H. Guo, E. Wu, M. Yan, and H. Zeng, ``High-Speed Mid-Infrared Single-Photon Upconversion Spectrometer," Laser Photonics Rev. \textbf{17}, 2300149 (2023).

\bibitem{Zhao2023NC} Y. Zhao, S. Kusama, Y. Furutani, W.-H. Huang, C.-W. Luo, and T. Fuji, ``High-speed scanless entire bandwidth mid-infrared chemical imaging," Nat . Commun. \textbf{14},  3929 (2023).

\bibitem{Israelsen2019LSA} N. M. Israelsen, C. R. Petersen, A. Barh, D. Jain, M. Jensen, G. Hannesschl\"{a}ger, P. Tidemand-Lichtenberg, C. Pedersen, A. Podoleanu, and O. Bang, ``Real-time high-resolution mid-infrared optical coherence tomography," Light  Sci. Appl. \textbf{8}, 11 (2019).

\bibitem{Junaid2019Optica} S. Junaid, S. C. Kumar, M. Mathez, M. Hermes, N. Stone, N. Shepherd, M. Ebrahim-Zadeh, P. Tidemand-Lichtenberg, and C. Pedersen, ``Video-rate, mid-infrared hyperspectral upconversion imaging," Optica \textbf{6}, 702-708 (2019).

\bibitem{Shahverdi2018OE} A. Shahverdi, Y. M. Sua, I. Dickson, M. Garikapati, and Y.-P. Huang, ``Mode selective up-conversion detection for LIDAR applications," Opt. Express \textbf{26}, 15914-15923 (2018).

\bibitem{Rehain2020NC} P. Rehain, Y. M. Sua, S. Zhu, I. Dickson, B. Muthuswamy, J. Ramanathan, A. Shahverdi, and Y.-P. Huang,  ``Noise-tolerant single photon sensitive three-dimensional imager," Nat. Commun. \textbf{11}, 921 (2020).
    
\bibitem{Ansari2018Optica} V. Ansari, J. M. Donohue, B. Brecht, and C. Silberhorn, ``Tailoring nonlinear processes for quantum optics with pulsed temporal-mode encodings," Optica \textbf{5}, 534-550 (2018).

\bibitem{Xia2015OL} H. Xia, G. Shentu, M. Shangguan, X. Xia, X. Jia, C. Wang, J. Zhang, J. S. Pelc, M. M. Fejer, Q. Zhang, X. Dou, and J.-W. Pan, ``Long-range micro-pulse aerosol lidar at 1.5 $\mu$m with an upconversion single-photon detector," Opt. Lett. \textbf{40}, 1579-1582 (2015).

\bibitem{Yue2022RS} W. Yue, T. Chen, W. Kong, X. Chen, G. Huang, and R. Shu, ``Eye-Safe Aerosol and Cloud Lidar Based on Free-Space Intracavity Upconversion Detection," Remote Sens. \textbf{14}, 2934 (2022).

\bibitem{Widarsson2020AO} M. Widarsson, M. Henriksson, P. Mutter, C. Canalias, V. Pasiskevicius, and F. Laurell, ``High resolution and sensitivity up-conversion mid-infrared photon-counting LIDAR," Appl. Optics \textbf{59}, 2365-2369 (2020).

\bibitem{Widarsson2022AO} M. Widarsson, M. Henriksson, L. Barrett, V. Pasiskevicius, and F. Laurell, ``Room temperature photon-counting lidar at 3 $\mu$m," Appl. Optics \textbf{61}, 884-889 (2022).

\bibitem{Wang2021PRL} B. Wang, M.-Y. Zheng, J.-J. Han, X. Huang, X.-P. Xie, F. Xu, Q. Zhang, and J.-W. Pan,``Non-Line-of-Sight Imaging with Picosecond Temporal Resolution," Phys. Rev. Lett. \textbf{127}, 053602 (2021).

\bibitem{Potma2021Optica} E. O. Potma, D. Knez, Y. Chen, Y. Davydova, A. Durkin, A. Fast, M. Balu, B. Norton-Baker, R. W. Martin, T. Baldacchini, and D. A. Fishman, ``Rapid chemically selective 3D imaging in the mid-infrared," Optica \textbf{8}, 995-1002 (2021).

\bibitem{Zhang2022PR} H. Zhang, S. Kumar, Y. M. Sua, S. Zhu, and Y.-P. Huang, ``Near-infrared 3D imaging with upconversion detection," Photonics Res. \textbf{10}, 2760-2767 (2022).
  
\bibitem{Fang2023LSA} J. Fang, K. Huang, E. Wu, M. Yan, and H. Zeng, ``Mid-infrared single-photon 3D imaging," Light  Sci. Appl. \textbf{12}, 144 (2023).  

\bibitem{ChenIEEE2023} Y. Chen, C. Jiang, Y. Liu, H. Su, X. Hu, and Y. Pan, ``A Compact Upconversion Single-photon Imager for Full-range and Accurate 3D Imaging," IEEE T. Instrum. Meas. \textbf{72}, 1502109 (2023).


\bibitem{Taschler2023LPR} P. T\"{a}schler, A. Forrer, M. Bertrand, F. Kapsalidis, M. Beck, and J. Faist, ``Asynchronous Upconversion Sampling of Frequency Modulated Combs," Laser Photonics Rev. \textbf{17}, 2200590 (2023).

\bibitem{Dong2018OL} X. Dong, X. Zhou, J. Kang, L. Chen, Z. Lei, C. Zhang, K. K. Y. Wong, and X. Zhang, ``Ultrafast time-stretch microscopy based on dual-comb asynchronous optical sampling," Opt. Lett. \textbf{43}, 2118-2121 (2018).  

\bibitem{Zhang2014OE} H. Zhang, H. Wei, X. Wu, H. Yang, and Y. Li, ``Absolute distance measurement by dual-comb nonlinear asynchronous optical sampling," Opt. Express  \textbf{22}, 6597-6604 (2014).

\bibitem{Ren2021IEEE} X. Ren, B. Xu, Q. Fei, Y. Liang, J. Ge, X. Wang, K. Huang, M. Yan, and H. Zeng, ``Single-Photon Counting Laser Ranging With Optical Frequency Combs," IEEE Photon. Technol. Lett. \textbf{33}, 27-30 (2021).

\bibitem{Huang2021PR} K. Huang, Y. Wang, J. Fang, W. Kang, Y. Sun, Y. Liang, Q. Hao, M. Yan, and H. Zeng, ``Mid-infrared photon counting and resolving via efficient frequency upconversion," Photonics Res. \textbf{9}, 259-265 (2021).
  
\bibitem{Yang2022SB} Y. Yang, Y. Jin, X. Xiang, T. Hao, W. Li, T. Liu, S. Zhang, N. Zhu, R. Dong, and M. Li, ``Single-photon microwave photonics," Sci. Bull. \textbf{67}, 700-706 (2022).

\bibitem{Yang2023FP} J. Yang, X. Zhao, L. Zhang, and Z. Zheng, ``Single-cavity dual-comb fiber lasers and their applications," Front. Phys. \textbf{10}, 1070284 (2023).
  
  
\end{thebibliography}
\end{document}